# New Theory of Superconductivity (Part I)

R. Riera and J.L. Marín

Departamento de Investigación en Física, Universidad de Sonora

Apdo. Post. 5-088, 83190 Hermosillo, Sonora, México.

A General Theory of Superconductivity with points of view differing from those of the BCS Theory is presented in two parts. In the first part, a general equation for the superconductivity is obtained; based on the stable pairing of two electrons bound by a phonon for any type of superconductor material. This equation comes from a self consistent field calculation with a screening which is temperature dependent; showing that the total energy of the electron pair is constant and the local energy of the paired electrons is equal to that of the phonon in the range $0\,K$ to $T_C$. A specific condition for the existence of the superconducting state is established allowing the prediction of the critical temperature. In the second part, the general equation is applied to low critical temperature superconductors where new results, in good agreement with experimental data, are calculated.

The BCS Microscopic Theory[1] was established in 1957 and within its more significant achievements are the hypothesis of the formation of correlated electron pairs mediated by the interchange of virtual phonons (Cooper pairs), and the formulation of an equation representing the law of dispersion of the electron pairs (Eqs. 2.49 and 2.50)[1]. However, the physical considerations for the pair formation taking into account the electron phonon interaction is not well defined because it incorporates as a result the sum of the electron – electron Coulomb repulsion and the electron phonon interaction as a constant and negative potential (Eq. 2.6)[1]; making the electron pair unstable and leaving a residual electric resistance.

On the other hand, in the same general equation the terms are misinterpreted and a dependence on the wave vector $\vec{k}$ for the total energy ($E_k$) is considered. Also, the local energy ($\varepsilon_k$) is a function of $\vec{k}$ and it is not equal to the phonon energy. Furthermore, equation 2.50 was not used for any calculation; likewise the energy gap (bonding energy) as a function of temperature. However, at $0\ K$ the ratio $2\varepsilon(0)/k_B T_C = 3.50$ (Eq. 3.50 Ref. 1) is fixed for all materials which does not discriminates any particular nature of the superconductor. BCS Theory does not establish a parameter characterizing each superconductor neither puts a condition to the existence of the superconducting state from which the critical temperature could be calculated.

Until now, there not exist a microscopic theory in which a physical discussion of the pair formation is done consistently neither obtaining a correct expression for the critical magnetic filed as a function of temperature. Above all, there is not theory using the fact that the energy of the local electron pair energy is equal to the phonon energy for the correctly calculation of the electron pair density of states from $0$ K to $T_C$.

In the framework of this new theory of superconductivity a solution for those problems is obtained by a logically and physically self consistent pathway.

In Part I, through a self consistent filed theory of the electron – electron Coulomb repulsion and the electron – phonon interaction, considering the temperature in the screening process, we establish as a condition for the superconducting state by identifying the coupling parameters of both interactions. From there, a general equation for the critical temperature can be derived. A conclusion obtained is that for a high $T_C$ superconductor it is necessary to have a big coupling of phonon – electron interaction. Also, we describe a model for the mechanism of electron pair formation based on the splitting and inversion of the symmetry point of the band when the material arrives to the critical temperature. We obtain a general equation for the superconducting state, similar to the law of dispersion of BCS theory but where the total energy of the paired electrons is conserved and where any local energy of the electron pair between 0 and $T_C$ only changes the internal state of the pair (bound energy). It is then clear then, that the superconductivity should occur when in the coupled electron pairs, the electron pair-phonon interaction and the electron-electron Coulomb repulsion are compensated and any change of energy between $0K \to T_C$ only produces a change in the internal state of the pairs, so that the total energy of pairs is conserved (see Fig. 1).

In Part II, we apply the general equation obtained in Part I to low temperature superconductors utilizing an *ad hoc* phononic theory, which is the case that BCS theory pretended to describe. In our work, we arrive, for the first time, to a specific expression for the bound energy as a function of temperature. We calculate correctly the density of states of the electron pairs and used it to get an equation for a temperature dependent critical magnetic field, needed in order to know the electrodynamical properties and Meissner effect. At the end, we include an expression for the specific heat as a function of temperature and compare it to experimental data for *Sn* and, also, calculate the jump at the critical temperature for eight superconductor materials that are in very good agreement with those reported in the literature.

Our purpose in this Part I, is to construct a general theory of superconductivity valid for any type of superconductor (high, intermediate or low temperature either). The general equation or law of dispersion of the superconducting state is obtained, considering that at the temperature of transition the electron pair-phonon interaction is annulled by the dispersion of Coulomb, and that is the only way of getting zero electric resistance. The starting points for the elaboration of our superconductivity theory are the following:

i) The microscopic carriers of the superconductivity are pairs of electrons bounded by phonons (a coupled electron pair). Because the electron-phonon interaction is the responsible of the electric resistance, it is necessary that two electrons form a paired state in order to the electron-electron Coulomb dispersion compensates the electron pair-phonon interaction and so eliminates the electric resistance;

ii) The transition to the superconducting state comes from a conducting phase where free electrons exist, allowing the formation of the pairs;

iii) The total energy of each pair is constant and it does not depend on the wave vector $k$ neither on the temperature $T$. When the temperature is lowered, starting from the critical temperature, the bounding energy of the pair increases to maintain the total energy constant and vice versa; i.e., an increase or decrease of temperature in the interval between 0 to $T_C$ only makes a change in the internal state of the pair in order to conserve the total energy, through a decrease or an increase of the phonon energy, respectively (see Fig. 1);

iv) The coupled electron pair is formed by three particles, two fermions and a boson, then it is nor a boson, neither a fermion, this is, it is not a problem of two bodies of identical particles but of three particles of different nature: they satisfy the Maxwell-Boltzmann statistics;

v) Because in the interval of temperature between $0K$ and $T_C$ the electrons in the pair and the phonon occupy the same energy states, the energy of the coupled electrons pair is equal to the energy of the phonon, which it depends linearly on their wave vector $q$ and on the temperature $T$. The superconductivity theory is sustained by a theory of phonons (harmonic, anharmonic, of low, intermediate and high temperature);

vi) The density of states of coupled electrons pairs in $T = T_C$ is the same and equal to the density of states of the free electron in the conductor state at the Fermi level. The

vii) density of states of the coupled electrons pairs between $0K$ and $T_C$ is similar to the density of states of the phonons;

vii) Upon arriving to the critical temperature, a splitting of the conduction band occurs due to a symmetry breaking in such a way that one electron sees the other as if its mass would be negative. This symmetry breaking leads also to an energy gap (energy of the bound of the electron pairs and gap for the free electrons) (centered on the Fermi level) between the coupled electrons (see Fig.2). A transformation of the electron energies of a three-dimensional reciprocal space of wave vectors $(k_x, k_y, k_z)$ to a bi-dimensional space of electron pair energies $(\ell(q), \varepsilon_0(T))$ has occurred (see Fig. 3).

Let us take a closer look to Fig. 2. The coupled electrons pair is constituted by two electrons that are slightly above and below of the Fermi level, with opposed moments $(k\downarrow, -k\uparrow)$ and they are changing permanently their state from the $(k\downarrow, -k\uparrow)$ to the $(k\downarrow +q, -k\uparrow -q)$ one and vice versa, by the emission (absorption) of a phonon of momentum $q$. Through this exchange the total energy reminds constant and the total momentum is zero. We also suppose that there exists a maximum binding energy of the pair, $2\varepsilon_0(T)$ when $T = 0K$, such that any energy beyond this limit leads to a pair breaking and to the lost of superconductivity state. This maximum energy depends only on the $T_C$, which is characteristic to each superconductor and establishes a forbidden zone for the not coupled electrons (energy gap).

Another property of the coupled electrons pair that it can be seen in Fig. 2 is that the electrons forming a pair belong to different bands. Therefore, one of the bands is inverted with respect to the Fermi level, behaving like a band of electrons of negative mass. The reason of this behavior comes from the fact that there not exist any band of the parabolic type (free electrons model) which passes simultaneously through the points $1(-k-q, \varepsilon)$, $1'(-k, \varepsilon + 2\varepsilon_0)$, $2(k+q, \varepsilon + 2\varepsilon_0)$ and $2'(k, \varepsilon)$, which it is due to the energies of electron pairs do not depend on the wave vector $k$; they only depend on $q$. In the same figure, the dashed lines correspond to the phonons and the solid lines to the electron transitions due to the emission and absorption of phonons. The electron in the positions 1' and 2 always emits phonons, while in the positions 1 and 2' absorbs phonons.

Notice that the states above $\varepsilon_2$ and below $\varepsilon_1$ are completely occupied, thus the simultaneous existence of conductivity and superconductivity is forbidden. We should take into account that only the conductivity by simple electrons can exist simultaneously with superconductor pairs, when there is a magnetic field and the electronic states of Landau are inside of the zone determined by $2\varepsilon_0$.

In Fig. 2 we can see that electron pairs are formed upon arriving to $T_C$ of the following form: $(k, -k, E(T_C))$ for the electrons $(k, \varepsilon(k))$, $(-k, \varepsilon(-k))$ and $(k+q, -k-q, E(T_C))$ for the electrons $(k+q, \varepsilon(k,q))$, $(-k-q, \varepsilon(-k-q))$. If there are electrons in $(k, \varepsilon(k))$ and $(-k, \varepsilon(-k))$ at $t = 0$, then in $(k+q, \varepsilon(k,q))$ and $(-k-q, \varepsilon(-k-q))$ there are unoccupied states (we will call them holes), in such a form that $\varepsilon(k) = \varepsilon(-k-q)$ and $\varepsilon(-k) = \varepsilon(k+q)$, this is to say, the electron energy is equal to the hole energy.

From Fig. 2 we can obtain the equations of the electronic bands as $\varepsilon(k) = -\hbar^2 k^2/2m + \varepsilon_F$ and $\varepsilon(-k) = \hbar^2 k^2/2m - \varepsilon_F$. Therefore $\varepsilon(k) - \varepsilon(-k) = 2\varepsilon(k) = \ell(k)$ and we can also find that $-(k+q) < -k_F < -k$ and $k < k_F < k+q$ and $k_F = k + q/2$.

We can derive the effect of lattice deformation on the electronic energy; simply we calculate the change in the total energy of the conductor due to the fact that the electrons (charged particles) can interact with the phonons (density waves in the charged lattice of ions). If this interaction is described by some interaction Hamiltonian $V^{ep}$, then the change in the energy of the conductor due to the interaction will be given, in second-order perturbation theory, by an expression of the following form:

$$\Delta\varepsilon = \sum_i \frac{\left|\langle f|V^{ep}|i\rangle\right|^2}{\varepsilon_f - \varepsilon_i}. \tag{1}$$

The excited state $|i\rangle$ corresponds to the electron in the state $|-k\rangle$ and the state $|f\rangle$ to the hole in the state $|-k-q\rangle$, then $\varepsilon_i = \varepsilon(-k) + \varepsilon_p(q)$, where $\varepsilon_p(q)$ is the phonon energy and $\varepsilon_f = \varepsilon(-k-q)$; however we already know that $\varepsilon(-k-q) = \varepsilon(k)$, therefore

$$\varepsilon_f - \varepsilon_i = \varepsilon(k) - \varepsilon(-k) - \varepsilon_p(q). \tag{2}$$

One such intermediate state is possible for every pair of occupied and unoccupied one-electron levels in the $|f\rangle$ state configuration. If we let $g_{-k,-k-q}$ be the matrix element of $V^{ep}$ between the $|f\rangle$ and $|i\rangle$ states, then the sum on $i$ is just a sum over all pairs of wave vector of occupied and unoccupied levels, and thus we have

$$\Delta\varepsilon = \sum_{-k,-k-q} n_k (1 - n_{-k-q}) \frac{\left|g_{-k,-k-q}\right|^2}{\varepsilon(k) - \varepsilon(-k) - \varepsilon_p(q)}; \tag{3}$$

now, considering that $v_{k,-k-q}^{eff-ep} = \partial^2 \Delta\varepsilon / \partial n_k \partial n_{-k-q}$, we obtain that

$$v_{k,-k-q}^{eff-ep} = -\left|g_{k,-k-q}\right|^2 \left(\frac{1}{\varepsilon(k) - \varepsilon(-k) - \varepsilon_p(q'')} + \frac{1}{\varepsilon(-k) - \varepsilon(k) - \varepsilon_p(q')}\right), \tag{4}$$

where $q'' = -q$ and $q' = q$; but $\varepsilon_p(-q) = \varepsilon_p(q)$, therefore we finally obtain that

$$v_{k,-k-q}^{eff-ep} = \left|g_{-k,-k-q}\right|^2 \left(\frac{2\varepsilon_p(q)}{\varepsilon_p^2(q) - \left[\varepsilon(k) - \varepsilon(-k)\right]^2}\right). \tag{5}$$

We argued that for many purposes the Fourier transform of the electron-electron Coulomb interaction should be screened by the electronic dielectric constant,

$$\frac{4\pi e^2}{q^2} \to \frac{4\pi e^2}{q^2 \xi^{el}} = \frac{4\pi e^2}{q^2 + k_0^2} \qquad q = -k - (-k - q), \tag{6}$$

where $\xi^{el}$ is the electronic dielectric constant that represents the effect of the other electrons in the screening of the interaction between a given pair. However, the ions also screen interactions, and we should have not use $\xi^{el}$, but the full dielectric constant $\xi(q)$ as $1/\xi(q) = \left(q^2/(q^2 + k_0^2)\right)\left(\omega^2/(\omega^2 - \omega^2(q))\right)$; then we find that the Eq. (6) should be replaced by

$$\frac{4\pi e^2}{q^2} \to \frac{4\pi e^2}{q^2 \xi(q)} = \frac{4\pi e^2}{q^2 + k_0^2}\left(1 + \frac{\omega^2(q)}{\omega^2 - \omega^2(q)}\right). \tag{7}$$

Thus the effect of the ions is to multiply the Eq. (6) by a correction factor that depends on frequency as well as on wave vector. As a result, the part of the effective electron-electron interaction mediated by the ions is retarded.

In order to use the Eq. (7) as an effective interaction between a pair of electrons, one needs to know how $\omega$ and $k$ depend on the quantum numbers of the pairs. We know that when the effective interaction is taken to have the frequency-independent form Eq. (6), then $q$ should be taken as the difference in the wave vectors of the two electronic levels. By analogy, when the effective interaction is frequency-dependent, we shall take $\omega$ as the difference in the angular frequencies (i.e., the energies divided by $\hbar$) of levels. Thus, given two electrons with wave vectors $-k$ and $k$ and energies $\varepsilon(k)$ and $\varepsilon(-k)$, we take their effective interaction to be [2,3]

$$V_{k,-k}^{eff-ee} = \frac{1}{V}\frac{4\pi e^2}{q^2 + k_0^2}\left(\frac{[\varepsilon(k) - \varepsilon(-k)]^2}{[\varepsilon(k) - \varepsilon(-k)]^2 - \varepsilon_p^2(q)}\right); \quad \omega = \frac{\varepsilon(k) - \varepsilon(-k)}{\hbar}; \quad \varepsilon_p(q) = \hbar\omega(q). \tag{8}$$

The Hamilton operator in the Hartree-Fock theory of the free electrons above the critical temperature can be written in the following form: $H_e = H_k + H_{ep}$, where $H_e$ is the total electron energy, $H_k$ is the kinetic energy of the free electron gas, and $H_{ep}$ is the electron-phonon interaction. Upon arriving to the critical temperature the electron pairs are formed and the Hamilton operator becomes in $H_p = H_{pk} + H_{ee} + H_{epe}$, where $H_p$ is total electron pair energy, $H_{pk}$ is the kinetic energy of the free pairs modified by a bounding energy that it characterizes the internal state of the electron pairs, $H_{ee}$ is the electron-electron Coulomb repulsion and $H_{epe} = H_{e_1 p} + H_{pe_2}$ is the electron pair-phonon interaction. $H_{ee}$ and $H_{epe}$ are related with the effective potentials $V_{-k,k}^{eff-ee}$ and $V_{-k,k}^{eff-ep}$ respectively, and both depend on the temperature, through the phonon energy.

Upon arriving to the critical temperature, it should fulfill that $V_{-k,k}^{eff-ee} + V_{-k,k}^{eff-ep} = 0$, so that the electron pairs remain free and its internal state is characterized by the bounding energy; then the electric resistance in the superconducting state doesn't exist. In the BCS theory $V_{-k,k}^{eff-ee} + V_{-k,k}^{eff-ep}$ represents a negative constant potential, which it makes unstable the electron pairs and does not eliminate the electric resistance. The BCS theory is a theory of the electron pairs unstable. So in order to $V_{-k,k}^{eff-ee} + V_{-k,k}^{eff-ep} = 0$ the coupling constants of electron-phonon interaction and electron-electron Coulomb repulsion should be equal, this is to say

$$\left|g_{-k,k}\right|^2 = \frac{1}{V} \frac{4\pi e^2}{q^2 + k_0^2} \frac{1}{2} \hbar \omega(q). \tag{9}$$

The condition given by the Eq. (9) and $\varepsilon(-k-q) = \varepsilon(k)$ convert the electron-phonon interaction in the electron pair-phonon interaction ($V_{-k,k}^{eff-ep} = V_{-k,k}^{eff-epe}$):

$$V_{-k,k}^{eff-epe} = \frac{1}{V} \frac{4\pi e^2}{q^2 + k_0^2} \left( \frac{\varepsilon_p^2(q)}{\varepsilon_p^2(q) - \ell^2(k)} \right), \quad V_{-k,k}^{eff-ee} = \frac{1}{V} \frac{4\pi e^2}{q^2 + k_0^2} \left( \frac{\ell^2(k)}{\ell^2(k) - \varepsilon_p^2(q)} \right). \tag{10}$$

As upon arriving to the critical temperature $T = T_C$,

$$V_{-k,k}^{eff-ee} + V_{-k,k}^{eff-epe} = \frac{1}{V} \frac{4\pi e^2}{q^2 + k_0^2} \left( \frac{\ell^2(k)}{\ell^2(k) - \varepsilon_p^2(q)} - \frac{\varepsilon_p^2(q)}{\ell^2(k) - \varepsilon_p^2(q)} \right) = 0, \qquad (11)$$

therefore $\ell^2(k) = \varepsilon_p^2(q)$, what it means that the electron pairs energy is equal to the phonons energy and in the case $T = T_C$ the phonon energy is maximum; then the electron pairs energy is maximum too and equal to $E^2 = \ell^2(q_{max})$, which it is constant.

As the Eq. (11) should be fulfilled in whole the interval of temperature of $T_C \to 0K$ and vice verse, taken into account that the phonon energy is a function of the temperature, then $q$ is also a function of the temperature and the electron pair energy is a function of the temperature too. Upon diminishing the temperature, starting from the critical temperature, the phonon energy or electron pair energy diminish, converting the differential of energy $\ell^2(q_{max}) - \varepsilon_p^2(q)$ or $E^2 - \ell^2(q)$ in energy of bond of the electron pairs $\Delta_0^2(T) = (2\varepsilon_0(T))^2$ in order to maintain the total energy $E$ constant. If we increased the temperature starting from $T = 0K$, then the phonon energy or the electron pair energy increase and the energy of bond diminishes in order to conserve $E$. Notice that $\varepsilon(k)$ and $\varepsilon_0(T)$ are the half of the electron pair energy and bond energy respectively, since they are defined $0K$ to $T_C$ (see Fig. 2). As conclusion we can write the following equations

$$\begin{aligned} E^2 - \ell^2(q) &= \Delta_0^2(T), & T_C \to 0K \\ E^2 - \Delta_0^2(T) &= \ell^2(q), & 0K \to T_C \end{aligned} ; \qquad (12)$$

combining both equations we arrived finally to the equation general of the superconductivity, which it represents the law of dispersion of the electron pairs:

$$E = \sqrt{\ell^2(q) + \Delta_0^2(T)} \quad \text{or} \quad E = \sqrt{\varepsilon_p^2(q) + \Delta_0^2(T)}; \qquad (13)$$

and considering that $q$ is a function of the temperature the Eq. (13) can be written of the following form:

$$E = \sqrt{\ell^2(T) + \Delta_0^2(T)} \quad \text{or} \quad E = \sqrt{\varepsilon_p^2(T) + \Delta_0^2(T)} . \tag{14}$$

The Eq. (14) is the general equation of the superconductivity and was obtained by BCS[1] and Bogoliubov[4], however they never used it neither they interpreted their terms correctly. In conclusion, upon arriving to the $T_C$ a breakup of the symmetry has occurred, with the splitting and reverting of the conduction band of the electrons and a canonical transformation of the 3D space in $(k_x, k_y, k_z)$ to the 2D space of energy $(\ell(T), \Delta_0(T))$ (see Fig. 3); starting from this figure the angle of dispersion can be calculated.

Until here we can conclude that the theory of superconductivity leads to a theory of the phonons (harmonic, anharmonic, of low, intermediate or high temperature).

Now, let us suppose there are only two physically distinguishable sources of scattering (for example, scattering by electron-phonon interaction and scattering by electron-electron Coulomb repulsion) as is the case. If the presence of one mechanism does not alter the way in which the other mechanism functions, then the total collision rate $W$ will be given by the sum of the collision rates due to the separate mechanisms $W = W^{ee} + W^{ep}$. In the relaxation-time approximation this immediately implies that $1/\tau = 1/\tau^{ee} + 1/\tau^{ep}$. If, addition, we assume a $k$-independent relaxation time for each mechanism, then, since resistivity is proportional to $1/\tau$, we will have $\rho = m/ne^2\tau = m/ne^2\tau^{ee} + m/ne^2\tau^{ep} = \rho^{ee} + \rho^{ep}$. This asserts that the resistivity in the presence of several distinct scattering mechanisms is simply the sum of the resistivities one would have if each mechanism were present alone.

The scattering rate $1/\tau$ depends on the square of the Fourier transform of the total interaction potential given by both interactions $V^{eff} = V^{eff-ee} + V^{eff-epe} = 0$. Therefore, the electric resistance is zero, which it is the condition for existence of the superconductivity.

The problems that previous theories of superconductivity have presented are due to an incorrect physical interpretation of this phenomenon and of the energies appearing in the general equation of superconductivity. Besides, those theories do not keep in mind that the electron pairs energies are equal to the phonon energies.

The relation $2\varepsilon_0(0) = 3.5 k_B T_C$ imposed in the BCS theory [1] forbids the existence of superconductivity of high critical temperature. This relation is important and it should be written in the following form: $2\varepsilon_0 = \lambda k_B T_C$, where $\lambda$ is our superconducting parameter, which it depends on each superconducting materials and it can be experimentally measured.

The isotopic effect in the low temperature superconductors, where the phonon theory is harmonic, can be described as $\omega^2(q) = \left(4\pi n_e Z^2 e^2 / \xi(q)\right)(1/M) = \left((1/\hbar)\lambda k_B T_C\right)^2$; then we can write $T_C = \gamma\left(1/\sqrt{M}\right)$, with $\gamma = \sqrt{4\pi n_e/\xi(q)}\left(Ze\hbar/\lambda k_B\right)$. The isotopic coefficient is 0.5 for these superconductors. It is necessary to notice that $\gamma$ is characteristic of each material. This permits that the isotopic effect can be screened according to $\gamma$, for some superconductors.

[1] J. Bardeen, L.N. Cooper and J.R. Schrieffer, Phys. Rev., **108**, 1175 (1957)

[2] H. Frohlich, Phys. Rev. **79**, 845 (1950)

[3] J. Bardeen and D. Pines, Phys. Rev. **99**, 1140 (1955)

[4] N. N. Bogoliubov, JETP **34**, 58 (1958)

Captions of figures

FIG. 1. A scheme of the energies that intervene in the general equation of the superconductivity, where $E$ is the total energy pairs, which it be conserved and $\ell = 2\varepsilon(k)$ is the electron pair energies equal to the phonon energies and $\Delta = 2\varepsilon_0(k)$ is the energy of bond, both vary of opposite form in order to maintain $E$ constant.

FIG. 2. We show a scheme of the way as the pairs are formed upon arriving to $T_C$ and the symmetry breaking leads to an energy gap centered on the Fermi level.

FIG. 3. We show the transformation of the electron energies of a three-dimensional reciprocal space of wave vectors $(k_x, k_y, k_z)$ to a bi-dimensional space of electron pair energies $(\ell(q), \varepsilon_0(T))$.

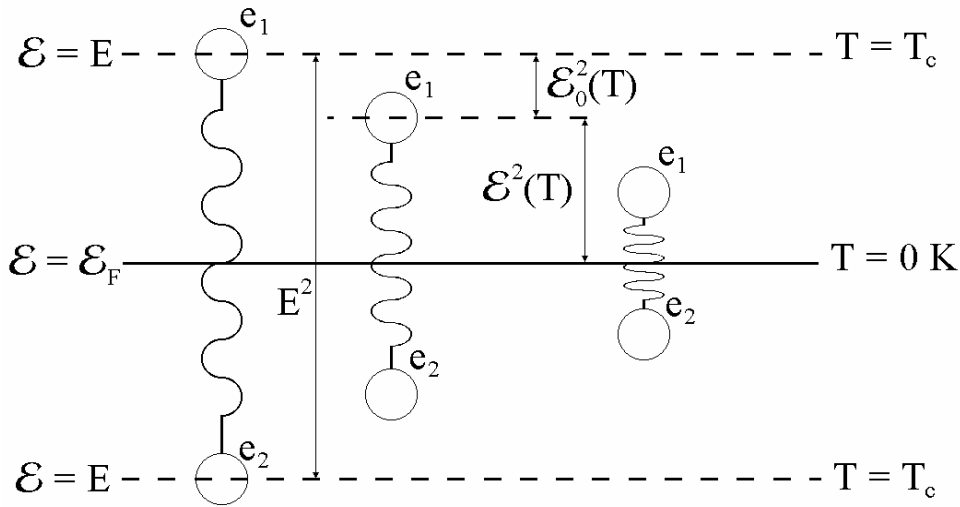

Fig.1

Fig.2

Fig.3